\documentclass[structabstract]{aa}  
\usepackage{graphicx}
\usepackage{txfonts}
\usepackage{natbib}
\begin{document}
   \title{AMBER/VLTI observations of 5 giant stars
   \thanks{Based on observations carried 
out at the European Southern Observatory (Paranal, Chile) under 
programme No. 082.D-0337.}}

%   \subtitle{I. Overviewing the $\kappa$-mechanism}

   \author{F. Cusano \inst{1, 2}
          \and
           C. Paladini\inst{3}
	  \and
          A. Richichi\inst{4,5}
	  \and
	  E. W. Guenther\inst{2}
	  \and 
	   B. Aringer \inst{6}
          \and \\
          K. Biazzo\inst{1}
	  \and
	  R. Molinaro\inst{1}
	  \and
	  L. Pasquini\inst{5}
	  \and
	  A. P. Hatzes\inst{2}}

   \institute{INAF-Osservatorio Astronomico di Capodimonte, Salita Moiariello16, I - 80131  Napoli, Italy
    \thanks{email: fcusano@na.astro.it}
\and Th\"uringer Landessternwarte Tautenburg, Sternwarte 5, D - 07778 Tautenburg, Germany 
%            \email{fcusano@na.astro.it}
\and  Institut f\"ur Astronomie der Universit\"at Wien, T\"urkenschanzstraße 17, A-1180 Wien, \"Osterreich
\and
National Astronomical Research Institute of Thailand,
191 Siriphanich Bldg., Huay Kaew Rd., Suthep, Muang
Chiang Mai 50200
Thailand
\and 
European Southern Observatory, Karl-Schwarzschildstr. 2, D-85748 Garching bei M\"unchen, Germany	   
  \and
	   INAF-Osservatorio Astronomico di Padova, Vicolo dell'Osservatorio 5, I - 35122 Padova, Italy }

   \date{Received; accepted}

% \abstract{}{}{}{}{} 
% 5 {} token are mandatory
 
  \abstract
   {While the search for exoplanets around main sequence stars more massive 
   than the Sun have found relatively few such objects, 
    surveys performed around giant stars have led to the discovery of 
   more than 30 new exoplanets. The interest in studying 
   planet hosting giant stars resides in  the possibility of  
    investigating planet formation around stars more massive  than the Sun. 
%    Theories state that  a correlation  between the mass of the 
%    hosting star and the one of the planet should exist; theories that have to be confirmed 
%    by  the observations. 
Masses of isolated 
   giant stars up to now were only  estimated from  evolutionary tracks, which led to different results
   depending on the physics considered. 
%and in the horizontal giant branch, tracks
%   of different masses all converge to the same HR diagram region.
    To calibrate the theory,   it is  therefore important   to measure a large number of 
   giant star diameters and masses as much as possible  independent of  physical models.  }
   {We aim in the determination of  diameters and effective temperatures  
   of 5 giant stars, one of which is known to host a planet. We used
    optical long baseline 
   interferometry with the aim of testing and constraining the theoretical models of giant stars.
    Future time-series spectroscopic observations of the same stars will allow the determination of  masses
by  combining    the asterosimological analysis and the interferometric diameter.}
  %  Combining the interferometrically-measured diameters with 
 %  asterosismological studies of  future time series spectroscopic observations,
 %  we will estimate  masses of giant stars without any dependence from models.}
   {AMBER/VLTI observations with the ATs were executed 
    in low resolution mode 
   on 5 giant stars.
    In order to measure  
    high accurate calibrated squared visibilities, a calibrator-star-calibrator observational sequence  was performed.  }
 {We measured the uniform disk and limb-darkened angular diameters of 4 giant stars. 
   The effective temperatures were also derived by combining 
the bolometric luminosities and the interferometric diameters. Lower effective 
temperatures were found when compared to spectroscopic measurements.   
The giant star HD12438 was found to have an unknown companion star at an  
    angular separation of $\sim$ 12 mas. 
    Radial velocity measurements present in the literature  confirm the presence of a companion with a very long orbital period (P $\sim$ 11.4 years).} 
   {}


   \maketitle
%
%________________________________________________________________

\section{Introduction}
The relation between the mass of hosting stars and planets has been investigated 
theoretically by several authors \citep{kornet2006,kennedy2008,
raymond2007}. According to these works there is a  dependence between 
the frequency of giant planets and the mass of the hosting star. 
Radial velocity (RV) surveys 
  clearly demonstrate that the frequency of Jupiter-mass planets around M-type stars
  is much lower than that of the more massive
  solar-like stars \citep{endl2006,johnson2007}.
%From an observational point of view the mass relation 
%between host star and planet has not been
%investigated properly due to the limitations 
%in detecting planet around   high mass  stars. 
Unfortunately, probing this relation for stars more massive than the Sun,
from an observational point of view, is quite difficult 
due to the limitations 
in detecting planets around high mass  stars.
Main sequence 
stars more massive than the Sun show  few 
and broad spectral lines, due to the higher effective  temperatures, gravity,
 and  rotation. Detecting  planets around these stars
 using the RV technique is very challenging, and up to now 
 only few planets have been discovered \citep[e. g., ][]{guenther2006, hartmann2010}. 
 The search for exoplanets 
%around massive star 
can be indeed performed on giant stars, which 
have more and sharper absorption lines respect to main sequence stars of similar 
masses, due to cooler photospheres, lower gravity, and slower rotation rate.
Up to now,  31 exoplanets are known orbiting giant stars (see Planet
Encyclopaedia\footnote{http://www.exoplanet.eu})
and certainly this number is destined to  increase in the near future
thanks to the  ongoing observational campaigns. In order to study the relation 
between the mass of the planet and the parent star, it is of fundamental 
importance to determine as accurately as possible the mass of the star. 
  Currently, masses of isolated giant stars are only determined by comparison with evolutionary models. 
  Although in the recent years the stellar evolutionary calculations have reached a highly sophisticated level,
there is still no convergence among the authors on the treatment of input physics
such as opacity and  metallicity. This can 
  lead to  different mass estimations for the same object.  
 % The different treatment of input physic
 % like for example the opacity table, reaction rates, metallicity etc can 
 % led to  different mass estimations for the same star. 
Therefore it
  is of fundamental importance to calibrate the theoretical models, 
  measuring physical parameters of a certain number of giant stars, using model independent
  techniques.

It is well established 
that giant stars show solar like pulsations, with typical periods longer than that of main sequence stars 
\citep{hekker2009,kallinger2010,bedding2010}. The asterosismological analysis of pulsating stars allows 
  one to derive the average density of a star, through the determination
  of the primary frequency splitting $\Delta\nu$ \citep[i. e. the frequency difference
between two harmonic modes characterized by different consecutive radial order $n$ and same 
angular degree $l$; ][]{kjeldsen1995}.  Combining
  this measure with the radius obtained using optical long baseline
  interferometry, it is then possible to derive the mass of the star.
  This technique has been successfully applied to the subgiant star $\beta$
  Hyi by \citet{north2007}, where the mass was determined with an error $<$ 3\%.
  \citet{hatzes2007} used RV measurements of the stellar oscillations
  in the planet hosting stars giant star $\beta$ Gem combined with the interferometrically 
  measured stellar radius to confirm that the star had a mass of $\approx$ 2 $M_\odot$.
  The frequency of maximum amplitude oscillations $\nu_{max} $ is also related to 
  the mass of the star, via  the radius and the temperature \citep[see][]{kjeldsen1995}. The detection 
  of this frequency is less observationally time consuming and gives also a good estimate of the mass.
  In principle, 
the mass and radius can be derived from a knowledge of $\Delta\nu$ and $\nu_{max} $, but
this is still not as accurate as using additional information on the
stellar radius. Unlike  main sequence stars, giant stars of similar effective temperatures
have a wide range of radii (10--60 $R_\odot$). An error of 20\%  in the radius
 corresponds to an error  of 60\% in the stellar mass, which is crucial for planet formation theory.
Thus asteroseismology alone, via the spacing of $p$-modes, does not 
provide  the mass of a star with the accuracy needed \citep[see][]{kjeldsen1995}.

  It is therefore essential to obtain good
 measurements of the angular diameter as an accurate stellar
 radius  is required by
  asteroseismic studies.  If one invests large amounts
 of telescope time to derive the oscillation spectrum, and then optical long baseline
  interferometry
 is unable to determine the radius (e.g. unresolved source), then
 the asteroseismic data will be of a limited use. 
 After  a large number of giant star diameters will be measured using interferometers, such
 as the  VLTI \citep[Very Large Telescope Interferometer;  ][]{haguenauer2008}, 
  then asterosismological observational campaigns can be performed on the most 
  interesting targets.  
  
  This paper is organized as follow: in Sect. 2 the observations 
  and the data reduction are presented; the determination of angular diameter and effective temperature 
by the UD-fit and LD-fit is presented in Sect. 3;  
Section 4 reports the discovery of the binary HD12438; 
results and conclusions are presented in Sect. 5.

 \section{Observations and data reduction}  
 \subsection{Description}
 Our project  started by observing interferometrically five giant stars with AMBER, 
 which is the near infrared VLTI instrument that  combines the light  from three
different telescopes \citep{petrov2007}. 
AMBER was used in combination with the Auxiliary Telescopes (ATs) adopting the telescope stations A0-K0-G1. The  configuration of the telescopes 
 is triangular, with two almost identical baselines  of 90.5 m (G1-K0 and A0-G1), and one  of 128 m (A0-K0).  
Observations were performed  in LR-mode (R=35) for a total of 20 hours spread over 9 different nights 
in the period between October and December 2008. A summary of the observations 
is presented in Table~\ref{table:obs}. 
The targets  chosen are listed in Table~\ref{table:targets}, together with some 
parameters from the literature.
The five  stars were selected from the \citet{dasilva2006} sample and were chosen 
in order to cover a wide range of metallicities, predicted masses, and radii.
This will allow us to   
 test theoretical models in different conditions.   
One of these stars, namely HD11977, is known to host a planet 
orbiting the star with a period of 711 days \citep{setiawan2005}. 
The predicted variability of these giant stars due to solar-like oscillations 
does not effect the interferometric observations, because
the radius variations due to the restoring of the $p$ force are just of
few kilometers at most. Additional sources of variability that could alter the interferometric observations are not 
known for our target stars.

\begin{table*}[t!]
\smallskip
\caption{ Observing log. All the observations were performed during the year 2008, using the ATs station A0-G1-K0,  DIT=50 ms and a CAL-SCI-CAL sequence. In the last columns the average statistical error in the H and K bands are given in percentage.}             
\label{table:obs}      
\centering          
\begin{tabular}{l c c c c c c }     % 7 columns 
\hline\hline       
                      % To combine 4 columns into a single one 
 Target  & Calibrator & Obs. date & N & $\sigma_{\rm stat}(H)$($\%$)  &  $\sigma_{\rm stat}(K)($\%$)$ \\
 
\hline   
                 
HD11977 &     HD5457,  HD24150,     &  7 Oct       &    2  & 5.2  &5.0  \\
        &                           &  22 Dec      &    1  & 6.1  &6.0   \\
HD12438 & HD10537, HD16897           & 5 Oct       &    1  & 6.4  &6.2   \\
        &                             & 6 Oct      &    2  & 5.3  &5.2   \\
        &                              & 20 Dec    &    1  & 5.4  &5.1  \\
HD23319 & HD22663, HD24150           &  7 Oct      &    1  & 3.3  &3.5  \\
        &                            &  10 Dec     &    1  & 4.2  &4.6   \\
HD27256 &     HD27442, HD28413       &    21 Dec   &    1  & 6.4  &6.1  \\
        &                             &   23 Dec   &    1  & 5.4  &5.3  \\
        &                             &   24 Dec   &    2  & 5.4  &5.1  \\
        &                             &  25 Dec    &    1  & 6.5  &6.6  \\
HD36848  &      HD33872, HD34642      &    21 Dec  &    1  & 7.2  &7.1  \\
         &                            &    22 Dec  &    1  & 8.2  &8.4  \\
         
 \hline                  
\hline
\end{tabular}
\end{table*}

\begin{table*}[t!]
\smallskip
\caption{Giant stars observed with AMBER.}             
\label{table:targets}      
\centering          
\begin{tabular}{l c c c c c c c}     % 7 columns 
\hline\hline       
                      % To combine 4 columns into a single one 
  ID       &  Ra         & Dec         & $V$      & $K$       &  Sp. Type   &  $\pi$    &  [Fe/H]\\
          &    (J2000)   & (J2000)     & (mag)  &  (mag)    &             &   (mas)  &   (dex)  \\
\hline
HD11977 &    01 54 56 & $-$67 38 50   &   4.70   &  $2.590\pm0.240$& G8.5 III & $14.91\pm0.16$ &  -0.21 \\
HD12438 &   02 01 15  &$-$30 00 07   &   5.35   &  $3.218\pm0.298$ & G5 III &   $11.08\pm0.29$ &  -0.61 \\
HD23319 &  03 42 50  & $-$37 18 49  &     4.60    & $2.639\pm0.274$ & K2.5 III & $17.70\pm0.22$ & 0.24  \\
HD27256 &  04 14 25 & $-$62 28 26 &   3.34   & $1.439\pm0.312$  & G8 II-III & $20.18\pm0.10$  &   0.07  \\
HD36848 &  05 32 51 & $-$38 30 48 &    5.46   & $2.804\pm0.268$  & K2 III  & $18.93\pm0.23$  &	0.21\\
\hline                  
\hline
\end{tabular}\\
{\tiny {Note that the $V$ magnitudes and the spectral types are taken from {\it Simbad}. 
The $K$ magnitudes are from 2MASS \citep{skrutskie2006}. The parallaxes are from the new reduction of Hipparcos data \citep{vanLeeuwen2007}.  The metallicities are from \citet{dasilva2006}}}.
\end{table*}

\subsection{Observation and data reduction}
In order to have reliable data, AMBER was used in combination with the fringe tracker.
 FINITO  is the VLTI fringe-tracker whose  purpose is
 to compensate the atmospheric turbulence effect on  two telescopes and to introduce
an additional difference in the optical difference path \citep{lebouquin2008}.
FINITO was used to track the fringes for all the objects except  HD12438 that was too faint.
A detector integration time (DIT) of 50 ms was used for each frame. Every  exposure
 consists of a cube of 1000 frames. 
%For each scientific and calibrator star 5 exposures were acquired during each observation.
Five exposures were acquired for each observation of a scientific (calibrator) star.
In order to obtain accurate measurements of the squared visibility, 
the observational sequence calibrator-star-calibrator (CAL-SCI-CAL) was adopted.
%we chosen an observational sequence calibrator-star-calibrator. 
This means
 that a calibrator star was observed shortly before and after the scientific target. 
The calibrator stars were chosen to be 
nearby  ($<10^{\rm o}$), and within
 a range  of {$\pm$ 0.5 mag} in the $K$ band with respect to the scientific target.
 The 9 calibrators (one calibrator star, HD24150, was used for two scientific targets) 
presented in Table~\ref{table:cal} were selected using the web-interface {\it CalVin}\footnote{http://www.eso.org/observing/etc/}.
 For each observational sequence, 1.50 hours were needed, including the internal calibration.

The data reduction was performed using the software {\it Amdlib}\footnote{http://www.jmmc.fr/data\_processing\_amber.htm}
 provided by the Jean-Marie Mariotti Center. 
Details of the AMBER data reduction
are explained  in  \citet{tatulli2007} and \citet{chelli2009}. 
In order to obtain the squared visibilities for each exposure, we
averaged  the frames  after a frame selection. The selection was performed keeping   20\% of the frames with the highest signal-to-noise
ratio (S/N). 
 This criterion was applied  for both the calibrators and  the science targets.

\subsection{ Visibility calibration}
\begin{figure}
\includegraphics[width=90.mm,height=85.mm]{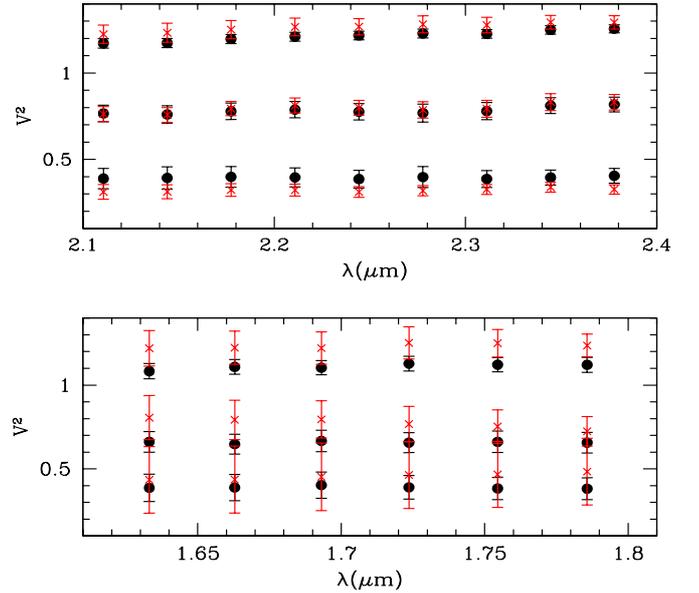}
\caption {Calibrated squared-visibilities of HD12438. The black dots are the 
calibrated visibilities using the calibrator HD16987, while the red crosses 
are those ones using the calibrator HD10537. More details are given in Sect. 2.3.} 
\label{img:calcal}
\end{figure}

The command line {\it amdlibDivide} in {\it Amdlib} was used to derive the calibrated squared visibility for each scientific 
exposure. 
 This command line derives calibrated visibility of the science targets  
using the following relationship: 
\begin{center}
\begin{equation}
\rm {V^2_{calS}=\frac{V^2_{obsS}}{V^2_{obsC}/V^2_{teoC}}}
\end{equation}
\end{center}
 where $\rm {V^2_{calS}}$ is the calibrated squared visibility of the science
 target, $\rm {V^2_{obsS}}$ is the observed squared visibility of the science target, $\rm {V^2_{obsC}}$ is the observed squared visibility of the calibrator star and  $\rm {V^2_{teoC}}$ the expected visibility of the calibrator star.
Each scientific exposure was calibrated with each  exposure of the two corresponding 
calibrators. 
The calibrated squared visibilities were then averaged. This procedure was performed for each observing block. 
Fig. \ref{img:calcal} shows  an example  of the squared visibilities 
of HD12438 (Obs. 5 Oct 2008) calibrated with HD19869 (dot symbols) and HD10537 (cross symbols). In order to make the plot more clear, the visibilities of the  baselines B2 and B3 where arbitrarly
shifted by $-$0.5 and +0.5, respectively. 
The final products  are   
 squared visibility measurements  in H and K band spectral channels.
 
\subsection{ Error estimation} 
  The final error on the calibrated squared visibility is given by two principal components:
 (i) a statistical error due to  the dispersion of the visibility in between the  single exposure; (ii)  
 a systematic error defined by the uncertainty in the calibrator angular diameter. The statistical error was computed taking the standard deviation
 relative to the averaged calibrated visibilities. The robustness of the choice of the standard deviation 
 was tested using a bootstrap method. 
 We combined the calibrated visibilities of one OB of HD27256 by using a bootstrap analysis \citep{efron1979}. 
 The resampling of the data was performed 500 times and the variance of the sample average was computed.
 This last resulted to be of the same order of the standard deviation for most of the data involved in the test.
 For this reason we kept the standard deviation of the average as the statistical error of   the calibrated visibilities.  
 
The effect of the calibrator angular diameter
uncertainty on the calibrated visibility was  computed 
analytically.  The computation of this  systematic error on the calibrated visibility
was performed by using for each target star the corresponding calibrators adopting the error on the diameter given in Table \ref{table:cal}. In the computations performed, we are also taken into account 
 the wavelength range and the  different baselines.
  The root mean square (rms) on the final calibrated-visibilities is of $\sim \%$ 2.4.

 The average error performed in the measure of the calibrated 
 visibilities is of $\sim 8\%$. In the best case we obtained 
 calibrated visibilities with 5.5$\%$ of accuracy (for HD27256), while in the worst case (for the star
 HD36848) we had 11$\%$ accuracy in one observing block.

\subsection{ Closure phase}

Together with the squared-visibility, another important observable that can be extracted by the AMBER  data is the closure phase (CP). 
The CP   allows one to improve the investigation of  the shape of a target. For example, 
the CP of an object with a circular-symmetric  distribution  of the light
 is always zero.
Reliable CP measurements  for the five giant stars were derived 
 by the AMBER data  using the software $Amdlib$. These CPs were calibrated
for instrumental effect by using the CPs measured for the calibrator stars. 
Further discussion on the CP   are presented in  Sect.~\ref{sec:bina}.
\section{Angular diameter and effective temperature determination}

 The  first step of our analysis was to estimate 
  the angular diameters of the target stars by fitting the AMBER data to a Uniform Disk (UD) 
analytical function. 
The UD is often used in the literature
to approximate the size of a single star \citep[e. g., ][]{dyck1998}.  
This approach 
can be very misleading for extended atmospheres \citep{jacob2002,paladini2009},
while it gives reasonable results in case of objects with an almost hydrostatic
atmosphere, like K-giants.

\begin{figure*}[t!]
\begin{tabular}{cc}
\multicolumn{1}{c}{\includegraphics[width=80.mm,height=70.mm]{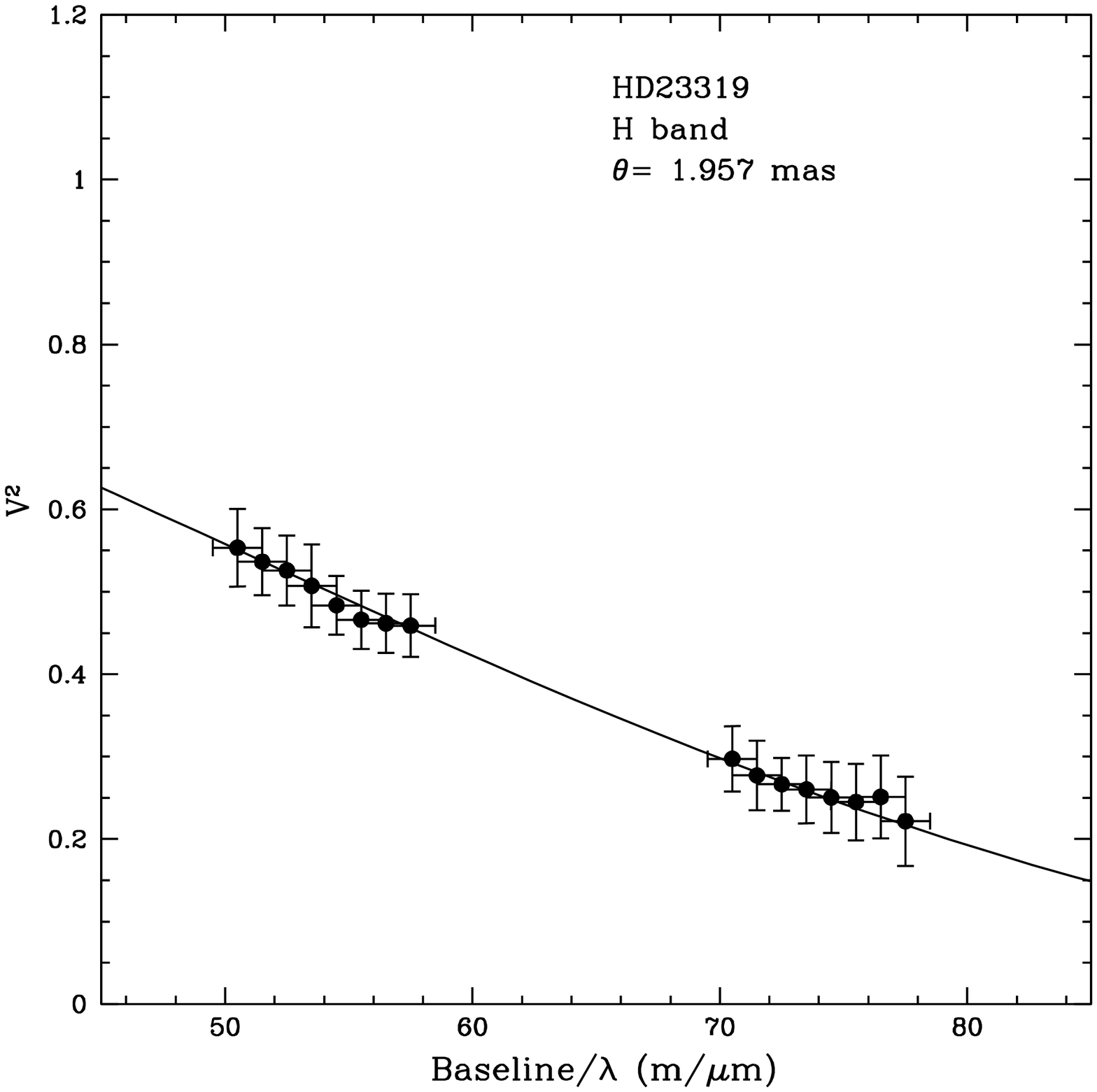}}&\multicolumn{1}{c}{\includegraphics[width=80.mm,height=70.mm]{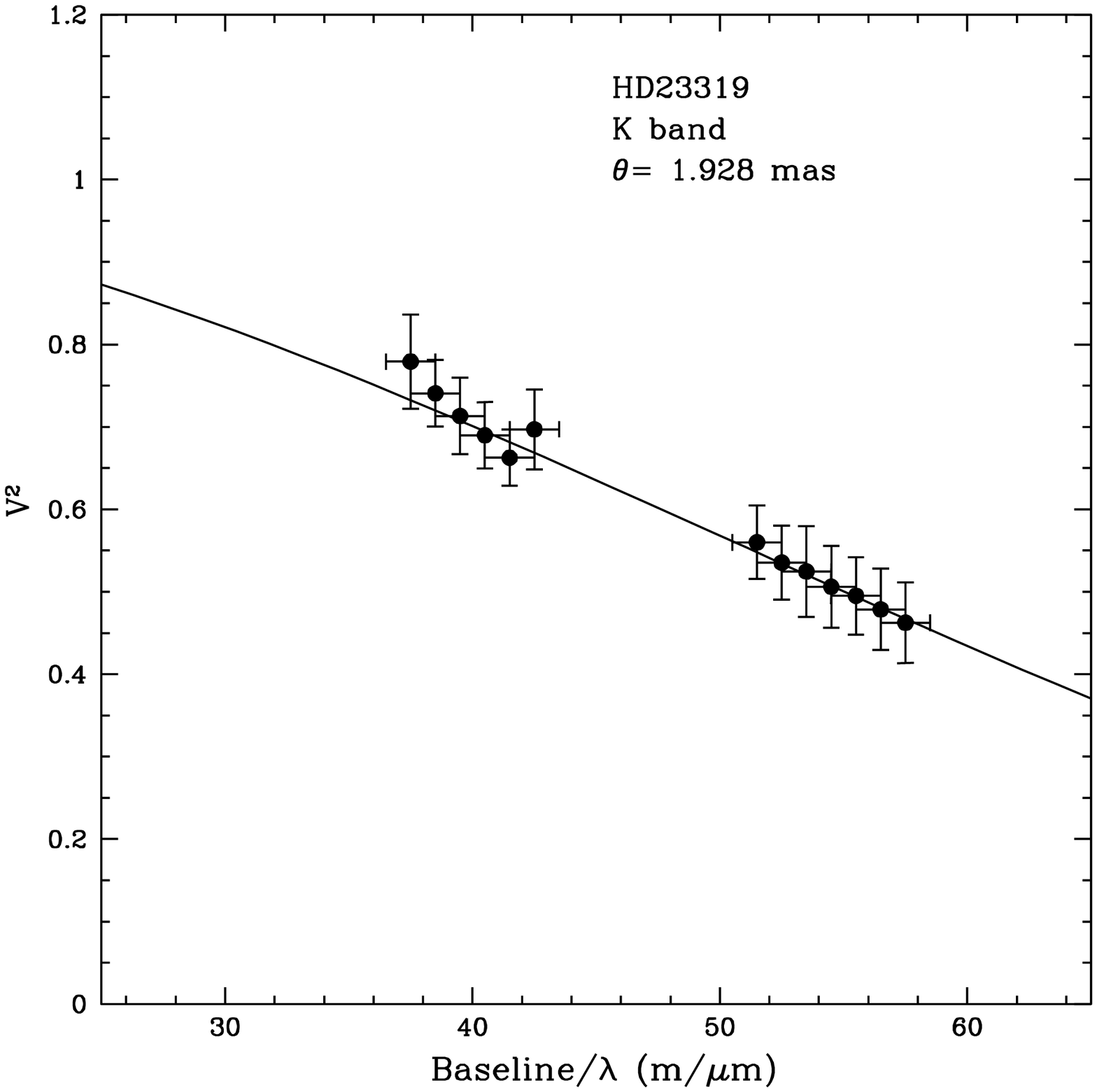}}\\
\end{tabular}
\caption{AMBER observed visibilities in the H (left) and K band (right) 
for HD23319. The line represent the UD model (cfr. Eq. \ref{equ:singd}) computed with best fitting angular diameters to the corresponding band.}
\label{img:visfits}
\end{figure*}

The angular diameters were derived fitting the squared-visibilities measured with AMBER to the model of a UD star:
\begin{equation}
V^2(\theta_{\rm UD}) = \left[\frac{2 {\rm J_1}(\frac{\pi {\rm B} \theta_{\rm UD}}{\lambda})}{\frac{\pi {\rm B} \theta_{\rm UD}}{\lambda}}\right]^2
\label{equ:singd}
\end{equation} 
where J$_1$ is the Bessel function of the first order, B is the projected baseline and $\theta$ is the stellar angular diameter. 
An example of the UD fit is shown in Fig.~\ref{img:visfits}. 
 Angular diameters of the scientific targets were derived by the UD-fit for 
each spectral channels in the $H$ and $K$ bands. 
 The errors are calculated by deriving the diameter at $\chi_{min}^2$+1 on both sides of the minimum $\chi^2$ and 
 determining the difference between the $\chi_{min}^2$ diameter and 
 $\chi_{min}^2$+1 diameter.  We thus computed the averaged diameters in the H and K bands, using the values 
derived in the spectral range 1.6-1.8 $\mu$m for the H band 
and 2.1-2.4  $\mu$m for the K band. The errors on the averaged H and K diameters
were  derived adding  the average error of the diameters calculated in the single spectral channel to the standard deviation. The UD angular
diameters are listed in Table \ref{tab:diamet}. 

%-------------------------------------------------------------
The limb-darkened diameters (LD) were obtained by fitting the data to 
equation
 \citep{hanbury1974}:
\begin{eqnarray}
 {\rm{V^2}}(\theta_{\rm LD})=\left(\frac{1-\mu_\lambda}{2} + \frac{\mu_\lambda}{3}\right)^{-1}\\
 \times \left[(1-\mu_\lambda)\frac{{\rm{J}_1(\frac{\pi{\rm{B}}\theta_{\rm LD}}{\lambda})}}{\frac{\pi{\rm{B}}\theta_{\rm LD}}{\lambda}}+ \mu_\lambda \left(\frac{\pi}{2}\right)^{1/2}+ \frac{{\rm{J}_{3/2}(\frac{\pi{\rm{B}}\theta_{\rm LD}}{\lambda})}}{\left(\frac{\pi{\rm{B}}\theta_{\rm LD}}{\lambda}\right)^{3/2}}\right]\nonumber
\end{eqnarray}

with $\mu_\lambda$ the linear LD coefficient. The latter was obtained
from \citet{claret1995} adopting the T$_{\rm{eff}}$ and log g values
 from  \citet{dasilva2006}. The errors on the LD
angular diameters were derived as explained above for  the UD angular diameters.
The  LD angular diameters  derived are listed in Table~\ref{tab:diamet}, together with  linear 
radii. These lasts  were calculated using the 
average  LD diameters and the parallaxes taken from \cite{vanLeeuwen2007}.

 Once the $\theta_{\rm LD}$ were determined interferometrically, the effective temperatures T$_{\rm{eff}}$ were calculated 
 by using the relationship:
 \begin{equation}
{\rm{T}}_{\rm{eff}}=\left(\frac{{4\rm{F}}_{\rm{bol}}}{\theta^2_{\rm{LD}}\sigma}\right)^{\frac{1}{4}}
\label{teff}
\end{equation}
where ${\rm{F}}_{\rm{bol}}$ is the bolometric flux and $\sigma$ is the Stefan-Boltzmann constant. The average of  the H and K band LD angular diameters  was used in Eq. \ref{teff}.
The  bolometric flux was calculated by applying  the bolometric corrections from \citet{pickles1998} to the de-reddened $V$ magnitude.  The A$_V$ for each star
 was derived by the E($V-K$) using the extinction curve reported by \citet{cardelli1989}. The E($V-K$) was calculated from the observed ($V-K$) and the intrinsic 
 ($V-K$)$_0$ using the spectral type 
reported in Table~\ref{table:targets} and the conversion table of \citet{pickles1998}. The final effective temperatures  are presented 
in Table~\ref{tab:diamet}.

%The approximately constancy of the diameters overall the two bands indicates %that the atmosphere can be considered
%compact, with no dependence of the disk edge    on the wavelength %\citep{baschek1991}.}

\begin{table}[t!]
\smallskip
\caption{Calibrator star informations. The parameters are taken from the  catalogs \citet{merand2005} and \citet{borde2002}.}             
\label{table:cal}      
\centering          
\begin{tabular}{l c c c c c }     % 7 columns 
\hline\hline       
                      % To combine 4 columns into a single one 
 Star  & $V$       & $K$    & Sp. Type & $\theta_{\rm UD}$ ($H$)  & $\theta_{\rm UD}$ ($K$) \\
       & (mag)    & (mag) &         & (mas)                 & (mas) \\
\hline                    
HD5457  &   5.46 & 2.98 & K2III     &  $1.29\pm0.02$      & $1.29\pm0.02$      \\
HD24150 &   6.76 & 2.56  & K5/M0III  &  $1.53\pm0.02$     & $1.55\pm0.02$      \\  
HD10537  & 5.26 & 2.95   & K0III   &  $1.29\pm0.02$   &  $1.30\pm0.02$ \\
HD16897  &7.40&  2.97   &  M0III  &  $1.27\pm0.02$        &  $1.28\pm0.02$     \\
HD22663  &4.59 & 2.04   &K1III   &  $1.89\pm0.02$         & $1.90\pm0.02$         \\
HD27442  &4.44 & 1.75   &K2IVa   &  $1.89\pm0.05$         & $1.90\pm0.05$           \\
HD28413  &5.95 & 2.41   &K4.5III  &  $1.85\pm0.05$        & $1.87\pm0.05$         \\
HD33872 &6.59 & 2.28   &K5III   &  $1.86\pm0.03$          & $1.88\pm0.03$        \\
 HD34642& 4.83 & 2.67  &K0IV   &  $1.49\pm0.02$           &$1.50\pm0.02$         \\
\hline                  
\hline
\end{tabular}
\end{table}

\begin{table*}
\smallskip
\centering
\caption{Diameters measured with AMBER for the five giants.   }
\label{tab:diamet}
\begin{tabular}{l c c c c c}
\hline
ID         & $\theta_{UD}(H)$&    $\theta_{UD}(K)$&$\theta_{LD}(H)$ & $\theta_{LD}(K)$ \\
           & (mas)       &         (mas)          &     (mas)       &(mas)               \\
\hline
\hline
HD11977   & $1.564\pm0.074^{stat} \pm0.024^{syst}$    &  $1.573\pm0.048^{stat} \pm0.033^{syst}$ &  $1.576\pm0.075^{stat} \pm0.025^{syst}$&$1.580\pm0.048^{stat} \pm0.033^{syst}$   \\   
HD12438*   & $1.0\pm0.1$    &  $1.0\pm0.1$ &                 &               \\ 
HD23319   & $1.957\pm0.026^{stat} \pm0.020^{syst}$    &  $1.928\pm0.044^{stat} \pm0.029^{syst}$ & $2.026\pm0.028^{stat} \pm0.021^{syst}$& $1.982\pm0.046^{stat} \pm0.028^{syst}$       \\
HD27256   & $2.559\pm0.025^{stat} \pm0.065^{syst}$    &  $2.557\pm0.028^{stat} \pm 0.071^{syst}$ & $2.639\pm0.027^{stat} \pm 0.071^{syst}$& $2.618\pm0.029^{stat} \pm 0.072^{syst}$    \\   
HD36848   &  $1.320\pm0.087^{stat} \pm0.018^{syst}$   &  $1.350\pm0.093^{stat} \pm0.025^{syst}$ & $1.363\pm0.090^{stat} \pm0.019^{syst}$&$1.386\pm0.095^{stat} \pm0.026^{syst}$   \\
\hline
\hline

\hline
\end{tabular}\\

*{\tiny {Note that for HD12438 we give the angular diameter of the primary, 
obtained by fitting the model of a resolved 
binary, as explained in the text.}}
\smallskip
\caption{ The linear radii and luminosities are derived by using  the Hipparcos
parallaxes reported in Table~\ref{table:targets}. }
\label{tab:radii}
\begin{tabular}{l c c c}
\hline
ID         & R$_{\rm{linear}}$     &L             & T$_{\rm{eff}}$  \\
           &   ($R_\odot$)         & ($L_\odot$)  &     ($K$)               \\
\hline
\hline
HD11977   & $11.38\pm0.65^{stat} \pm0.32^{syst}$     & $45.3\pm8.4$   &  $4445\pm125$     \\   
HD12438   & $9.71\pm1.34$&  $48.0\pm19.3$  &  $4884\pm250$    \\ 
HD23319   &$12.18\pm0.36^{stat} \pm0.26^{syst}$   & $45.2\pm4.5$   &  $4294\pm58$        \\
HD27256   & $14.01\pm0.22^{stat} \pm0.54^{syst}$    & $91.6\pm5.8$  &  $4777\pm36$    \\   
HD36848   &  $7.81\pm0.75^{stat} \pm0.20^{syst}$ & $17.4\pm5.1$   &  $4223\pm209$     \\
\hline
\hline

\hline
\end{tabular}

\end{table*}

\section{Binary discovery: HD12438}
\label{sec:bina}
\begin{figure*}
\begin{center}
\includegraphics[width=180.mm,height=90.mm]{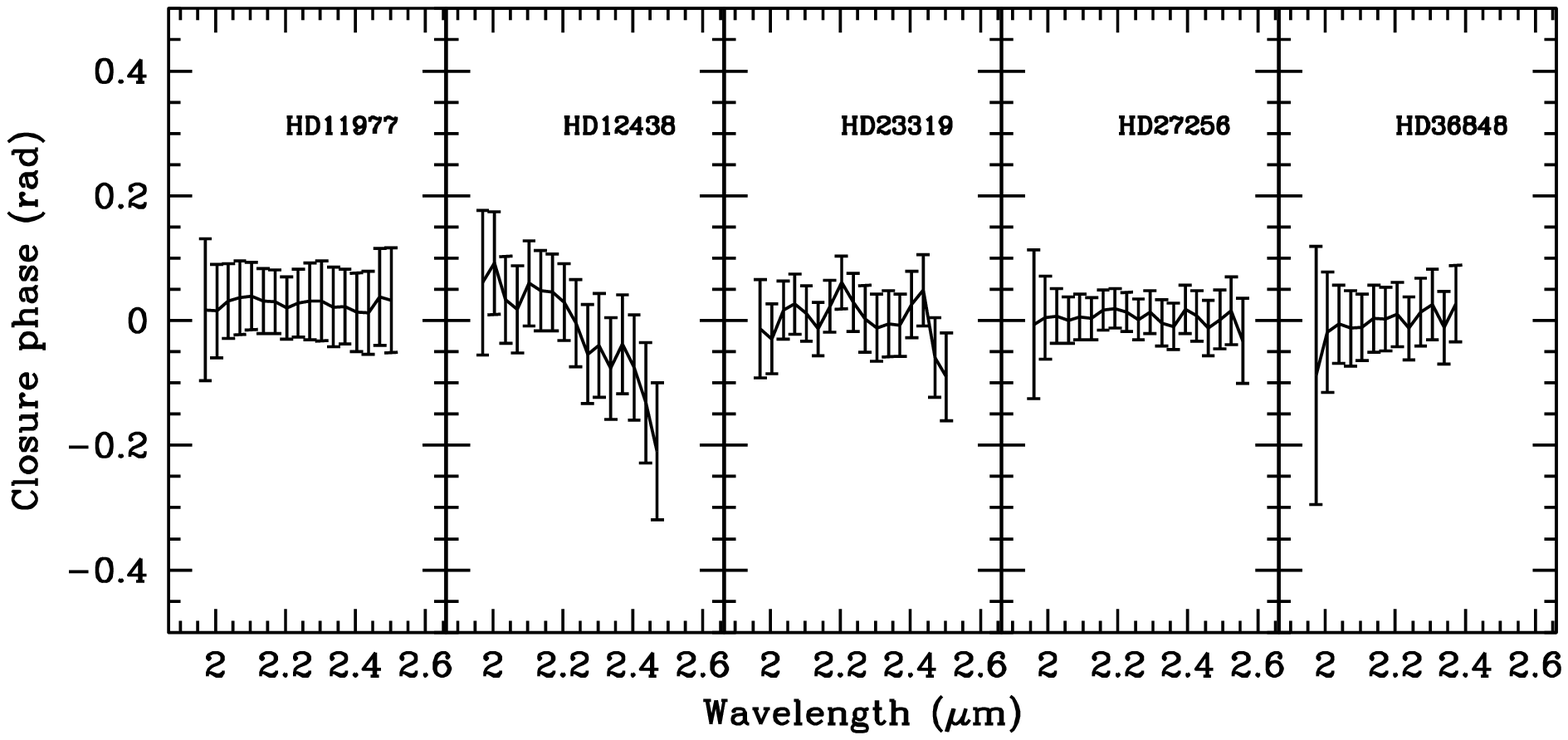}
\caption{Closure phase in the $K$ band for the 5 giant stars observed with AMBER. HD12438, that we found to be a binary, 
has a closure phase different from zero for wavelengths $>$ 2.1 $\mu$m. The other giant stars have  closure phase close to zero, as expected.}
\label{img:closphas} 
\end{center}
\end{figure*}

We were able to fit the angular diameters in each spectral channel with
 $<$ 5.5\% accuracy  for all the stars, 
with the exception of HD12438.
The quality of the data for this star 
was the same as for the other scientific targets. We speculated at this point that this star could have a companion; in fact  
a single star model was not able to reproduce 
the visibilities. A clue in this direction was given by the fact that the measured visibilities on the baselines G1-K0 and A0-G1, that have almost the same length, 
but different orientation, were different. In order to confirm our hypothesis we developed at this point a software to fit the observed visibilities to a model of a pair of resolved stars.  HD12438 
was observed 4 times with AMBER (see Table~\ref{table:obs}). More precisely, one observation was performed  on the  5th of October,
 two on the 6th of October and one  on the 20th of December, in 2008.   
 The software  calculates the angular separation between the components of the binary, the position 
 angle, the flux ratio, and the angular diameters of the two components. The  squared visibilities in this case are given by the equation:
 \begin{equation}
{\rm{V^2}} = \frac{{\rm{J}}_1(\theta_1) + {\rm{J}}_2(\theta_2) {\rm{f^2}} + 2 {\rm{J}}_1(\theta_1) {\rm{J}}_2(\theta_2) {\rm{f}}\cos(2 \pi\frac{{\rm{\bf{B\cdot\rho}}}}{\lambda})}{(1 + {\rm{f}})^2} 
\label{equ:bin}
 \end{equation} 
 where f is the flux ratio, J$_1$ and J$_2$ are the Bessel functions corresponding to the single binary components with angular diameters $\theta_1$ and $\theta_2$,
 {\bf{B}} is the baseline vector, and {\bf{$\rho$}} is the separation vector.  The whole set of data of this star fits
    to the Eq. \ref{equ:bin} better than  to the relationship valid of a single star with an UD. 
Fitting the observed visibilities  acquired at four different epochs to the Eq. \ref{equ:bin}, 
 we  estimated the separation angle between the components,  which is $\rho=12.0\pm4.0$ mas. 
The position angle of the angular separation is barely changing for observations separated by 2.5 months (6 October -
20 December 2008), indicating a  long orbital period. The position angle has an average value of $120^{\rm o}\pm20^{\rm o}$. 
\begin{figure*}[t!]
\begin{tabular}{cc}
\multicolumn{1}{c}{\includegraphics[width=80.mm,height=70.mm]{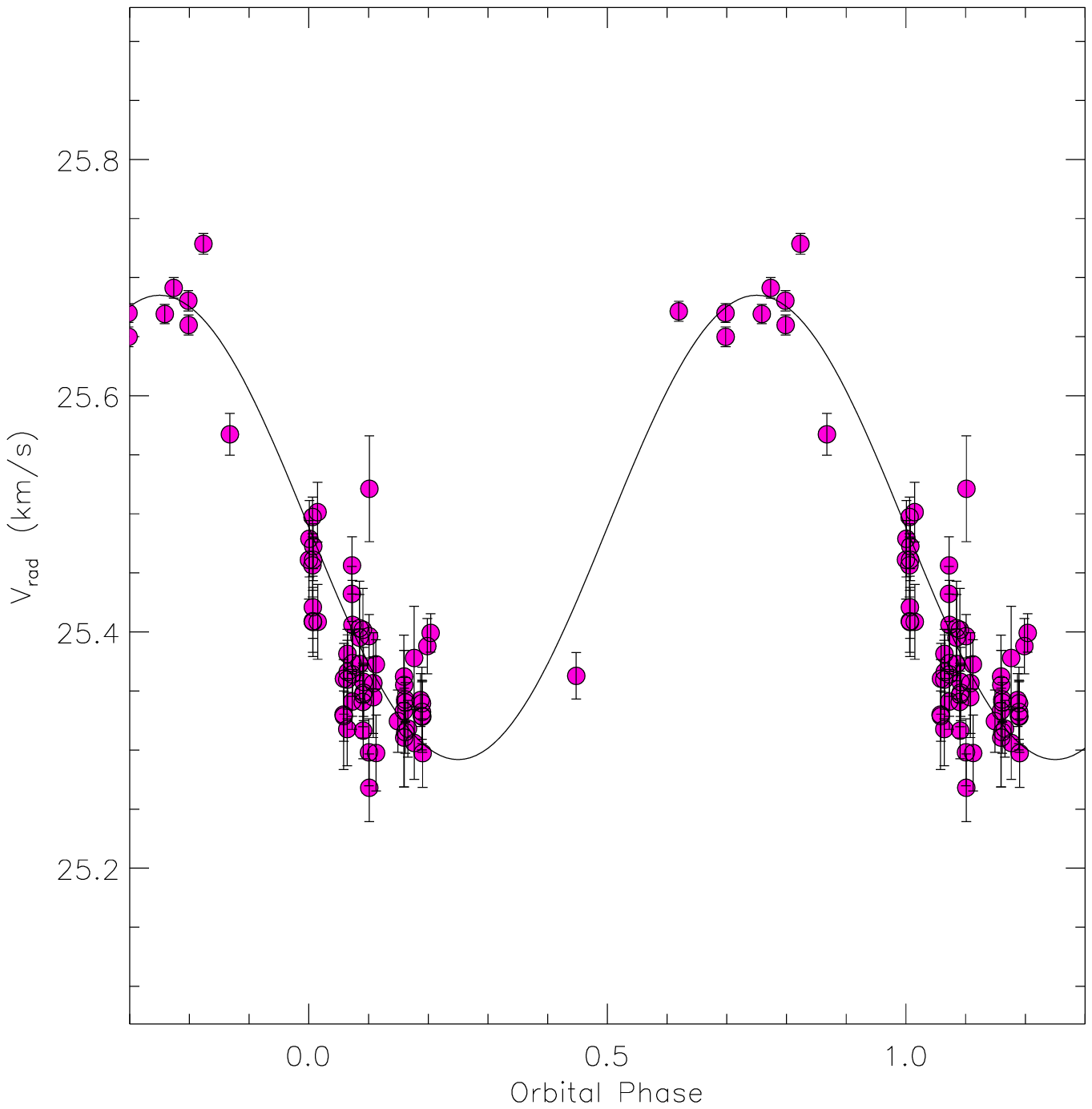}}&\multicolumn{1}{c}{\includegraphics[width=80.mm,height=70.mm]{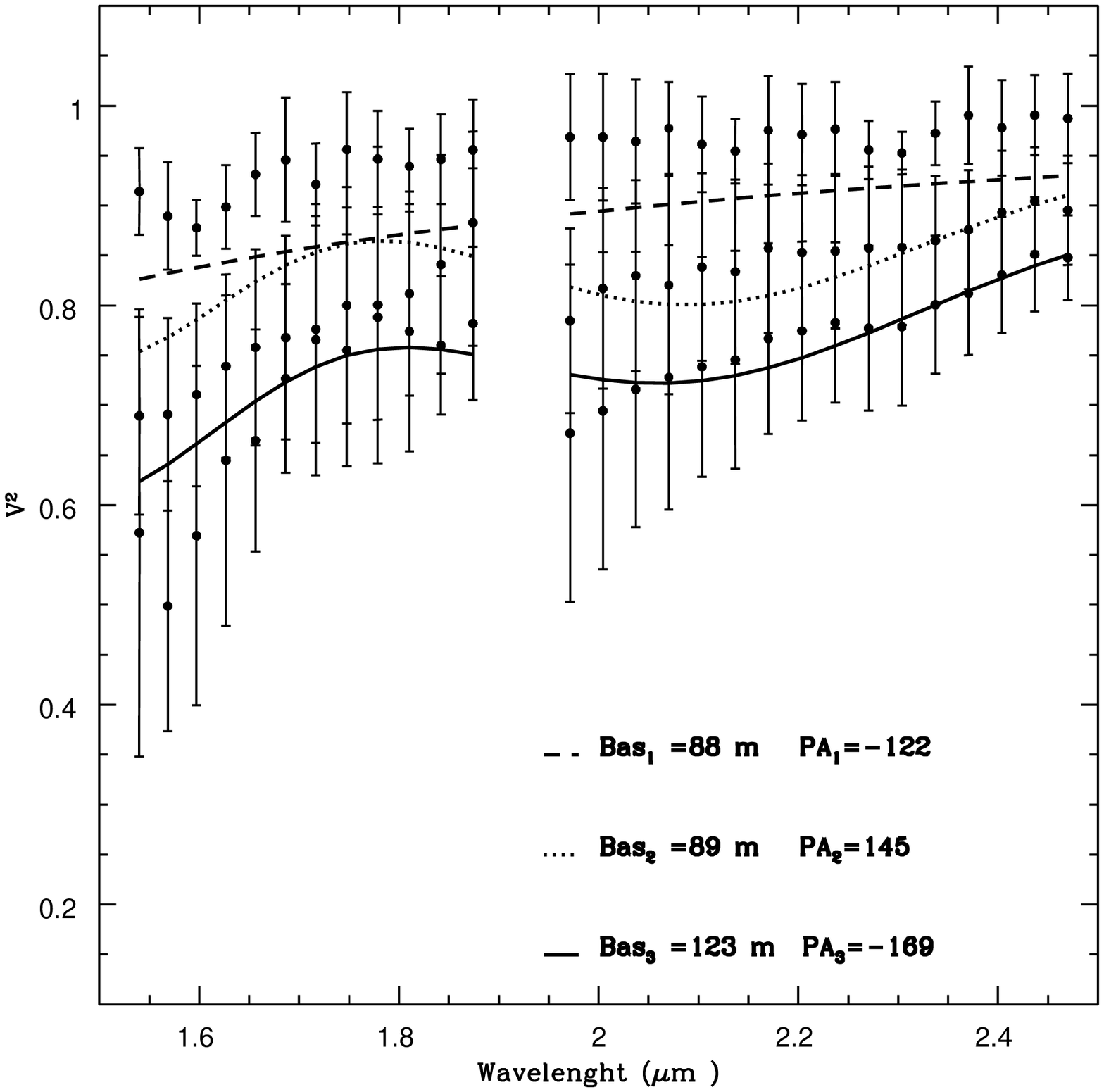}}\\
\end{tabular}
\caption{$Left$: RV data of HD12438 obtained with FEROS and HARPS during more than 9 years. 
The line represent the orbit with lower rms obtained 
for a period of 11.4 years fixing the eccentricity to e=0. 
The high scatter around this line is mainly due to stellar oscillations.
 $Right$: Squared visibilities  
of HD12438 observed  the  6th of October 2008. The lines are the visibilities of a binary (see Eq. \ref{equ:bin}), for the baselines of the observation, obtained with a   
 separation of the components of  $\sim$ 12 mas and a  position 
angle of 120$^{\rm o}$.}
\label{hd12438}
\end{figure*}

The primary star is resolved with the baselines chosen. The angular 
 diameters of the primary and secondary are respectively  of $1.0\pm0.1$ mas and $\leq$ 0.3 mas.   
 The flux ratio ratio is  $\sim$0.028 both in  the $H$ and $K$ bands. 
The calibrated closure phase also indicated that the object is not centro-symmetric, as it is different from zero
 to certain wavelengths. The comparison
with the closure phases of the other 4 giant stars, presented in Fig.~\ref{img:closphas}, clearly shows that HD12438 is not a single star.  

 Further confidence of the binarity of this object was given by the spectroscopy 
 present in the literature. HD12438 was investigated spectroscopically  as a part of planet search programs by 
  \citet{setiawan2004} and Hatzes (private communication) using the FEROS$@$ESO and HARPS$@$ESO spectrographs, respectively. 
 The former  observed this object for 5 years obtaining 61 spectra 
 and finding a small linear trend in the radial velocities (RVs)  with the time.  The latter obtained 9 spectra   
 distributed over  2.8 years. 
 The whole FEROS+HARPS spectroscopic observations cover a period of more than 9 years. We analyzed the RVs 
 obtained with these spectrographs, correcting the values for the offset of 0.180 km/s obtained by \citet{demedeiros2009}.
 The result is shown in Fig. \ref{hd12438}, where an orbital period of 11.4 years was found. A detailed discussion of the 
 spectroscopic orbital parameters is far from the purpose of this work, where the RV data were used by us only to confirm the binarity 
 of HD12438. 
 Using a mass value for the primary component of M$_1$=$1.02\pm0.19$ $M_\odot$ \citep{dasilva2006} 
  and assuming a mass of the secondary of M$_2\sim0.5$ $M_\odot$ deduced from the flux ratio, 
  the angular semi-mayor axis for a minimum 
  orbital period of 11 years (at a distance of 52.8 pc, see
  Table~\ref{table:targets}) is $>$ 70 mas. This means that the binary is highly inclined toward the line of sight. 
  This  also explains  the small RV amplitude variations observed.
 
As previously mentioned, by fitting   Eq. \ref{equ:bin} we estimated the angular diameter of the primary component. This value is in agreement with the 
estimate by \citet{dasilva2006}. The temperature of the primary was derived by combining the averaged UD angular diameter and the bolometric luminosity.

\section{Discussion and conclusions}
\begin{figure*}[t!]
\begin{tabular}{cc}
\multicolumn{1}{c}{\includegraphics[width=80.mm,height=70.mm]{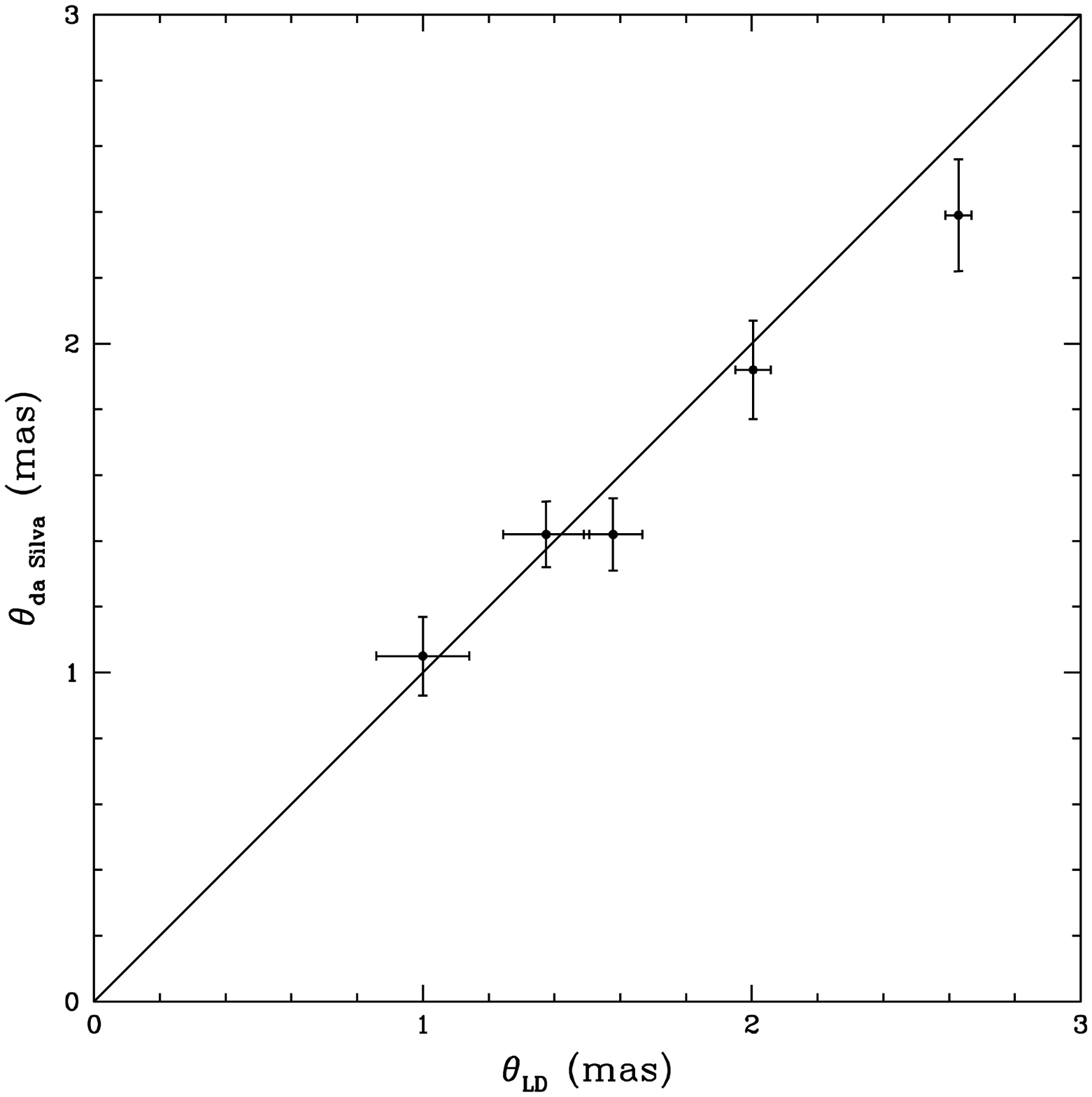}}&\multicolumn{1}{c}{\includegraphics[width=80.mm,height=70.mm]{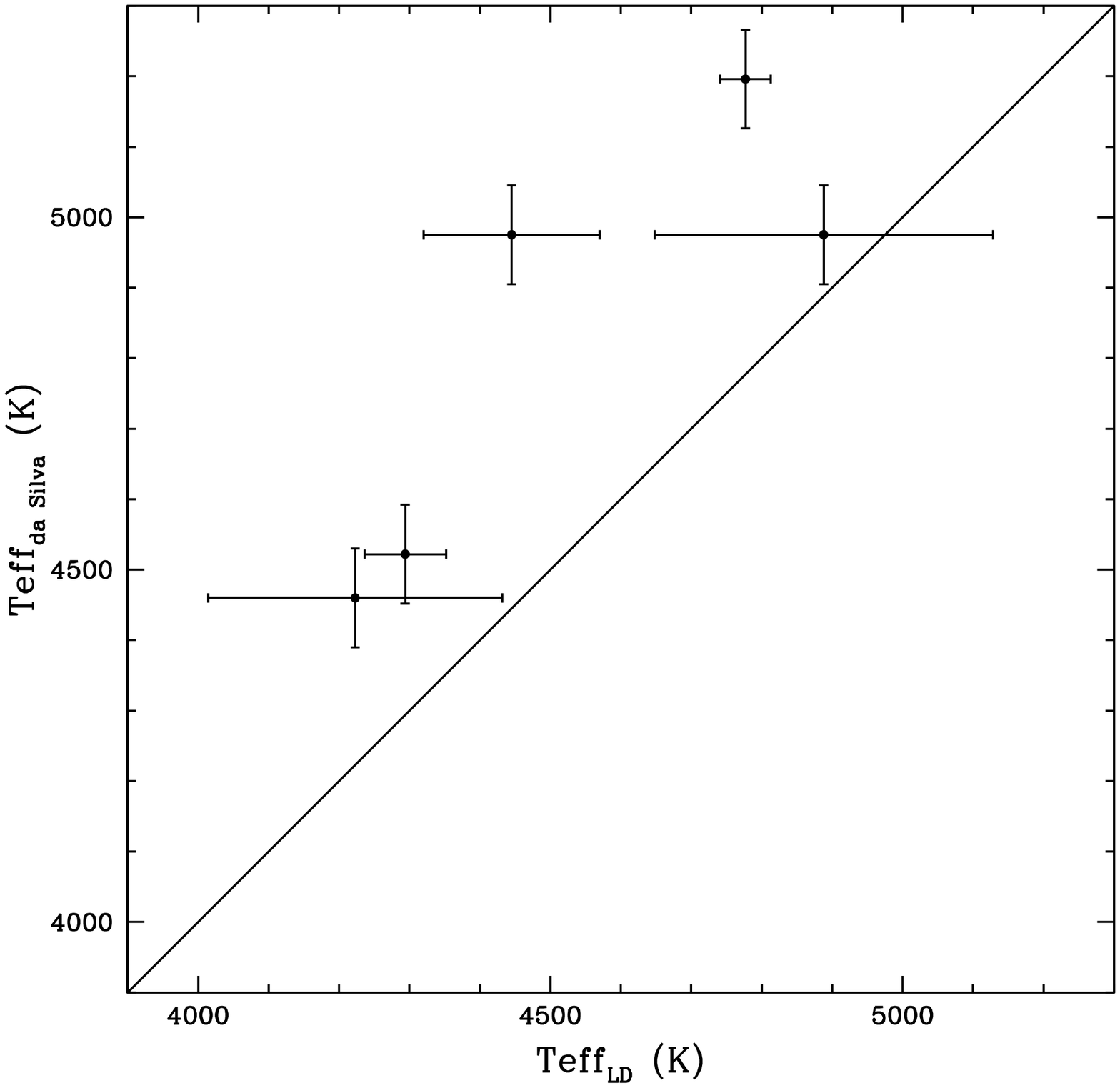}}\\
\end{tabular}
\caption{ $Left$: Comparison between interferometric-measured LD angular  diameters and 
model precition by \citet{dasilva2006}. The LD diameters were obtained averaging
the LD diameters in the H and K bands.  
 $Right$: Comparison between the effective temperatures obtained using the  Eq.\ref{teff}
 and the spectroscopic temperature derived by \citet{dasilva2006}. The 
 interferometric derived temperature are systematically lower than the 
 spectroscopic ones.}
\label{confr}
\end{figure*}

 In this work we  derived  angular  diameters of five field giant stars, selected from the sample of \citet{dasilva2006}.
In particular, using the AMBER$@$VLTI instrument combined with the fringe tracker FINITO, 
 we were able to measure through UD and LD-fitting the angular 
diameters of HD23319   and HD27256 with  accuracy of $3.2\%$ and $4.1\%$,
respectively  when both the statistical and systematic error are taken into account. For HD11977 and HD36848 the relative error of the angular diameter was of $6.2\%$ and $9.8\%$, respectively.
   The visibilities of the giant star HD12438  were not reproducible by a single star model. We detected the companion star 
   by interferometry, and we confirm and assess the period  from radial velocity measurements. 
Fig. \ref{confr} shows the comparison between the LD angular diameters averaged overall the H and K bands and the predicted  angular diameters
 derived by \citet{dasilva2006}. The two set of values are consistent within 1.5 $\sigma$. The difference between UD and LD diameters for our sample of stars is
  smaller than $< 0.8\%$, and in the case of HD11977 $\sim 0.15\%$.

Combining the LD angular diameters and the bolometric fluxes, we were able to derive the effective temperatures, which, as shown in Fig. \ref{confr}, 
are always 
lower than the values coming from spectroscopic analysis \citep[see ][]{dasilva2006}. This is interesting as  
 as far as the ongoing debate on metallicity of giant stars is concerned. 
 The \citet{dasilva2006} data analysis is at the basis of the results which were used by
 \citet{pasquini2007} to show that giants hosting planets are not preferentially metal rich, in contrast to the main
 sequence stars. 
 One possibility to explain the discrepancy with
 main sequence stars is that  metallicities of giant stars are systematically too low \citep[see, e.g., ][]{santos2008}. 
 In fact, the iron abundance  depends on  other parameters  such as  T$_{\rm eff}$, log g, and microturbulence.
 A  higher T$_{\rm eff}$
corresponds to a higher [Fe/H], if the log g  and the microturbulent velocity are kept constant.
 Our LD interferometric results (LD-fitting)  show lower temperatures for giant stars when compared to \citet{dasilva2006}, and are indeed in  the direction of
  a lower  estimate 
of [Fe/H], supporting the idea of  \citet{pasquini2007}.  
Similar result was reported by \citet{biazzo2007}, which found on
average a 
lower T$_{\rm eff}$ compared to \citet{dasilva2006}.

The case of HD12438 points out that high inclined stellar binary systems
which can be      
wrongly identified as planet hosting stars are not so rare. HD12438 shows
that optical long baseline interferometry  is a useful 
tool to detect such systems. The binarity of the system, in fact, can be
only detected  if the flux ratio between the primary star 
and the companion is  much higher  than typical  value for star$/$planet.

%%The angular diameters derived by the UD-fit averaged on the H and K band, 
%are consistent in 3 $\sigma$ with the angular diameters derived by fitting visibility profiles generated by COMARCS  model atmosphere 
%and   with  the estimates given by \citet{dasilva2006}. 
% 

% The temperatures derived  by combining the UD diameters and the bolometric fluxes 
% are always lower than the one 
%coming from the spectroscopic analysis. A similar results 
% was found recently by \citet{baines2010}.
% This difference is due to the different  properties of the two measurements.
 
% The spectroscopic temperatures are derived generally by  using  the neutrum and ioneized iron lines. 
%The temperature derived using this method gives an average 
%value of the part of the atmosphere where the iron lines are formed. The AMBER observations instead measured the temperatures
%of the upper part of the atmosphere  visible in the near infrared. 

%At this stage, the visibility profiles derived  by  the COMARCS  model atmosphere are too coarse and simple to have an 
%accurate determination  of the parameters of the stars.

%In all cases the log g from the profile  underestimates the spectroscopy. The rise of the temperature in the model has the effect of increasing   
%the log g. This can explain the systematic higher temperature derived by the profiles.

Future high accurate spectroscopic time-series observations of these giant stars will allow to determine 
the masses by combining the asterosismologic analysis with the interferometric diameter. In the case 
of HD12438 spectroscopic observation will also constrain the  binary orbit.

\begin{acknowledgements}
 We thank the anonymous referee for the useful comments and suggestions. 
 This research has made use of the  \texttt{AMBER data reduction package} of the
Jean-Marie Mariotti Center\footnote{Available at http://www.jmmc.fr/amberdrs}.   
C. Paladini acknowledge the support of the Project  P19503-N16
of the Austrian Science Fund (FWF).
\end{acknowledgements}

\bibliographystyle{aa}
\bibliography{mybib}

%\begin{thebibliography}{}

\end{document}